\documentclass[preprint]{aastex}
\usepackage[T1]{fontenc}
\usepackage[utf8]{inputenc}
\usepackage{epsfig}
\usepackage{amsmath}
\usepackage{pifont}
\usepackage{url}
\usepackage{graphicx}
\usepackage[normalem]{ulem}

\shorttitle{9577 and 9633 DIBs}

\shortauthors{Galazutdinov et al.}

\begin{document}

\title{Diffuse bands 9577 and 9633 -- relations to other interstellar features}

\author{G.A.~Galazutdinov}
\affil{Instituto de Astronomia, Universidad Catolica del Norte
    Av. Angamos 0610, Antofagasta, Chile\\
Pulkovo Observatory, Pulkovskoe Shosse 65, Saint-Petersburg 196140, Russia\\
Special Astrophysical Observatory of the Russian AS, Nizhnij
Arkhyz 369167, Russia }
\email{runizag@gmail.com}

\author{G. Valyavin}
\affil{Special Astrophysical Observatory of the Russian AS, Nizhnij Arkhyz 369167, Russia}

\author{N.R.~Ikhsanov}
\affil{Pulkovo Observatory, Pulkovskoe Shosse 65, Saint-Petersburg 196140, Russia}

\author{J.~Kre{\l}owski}
\affil{Materials Spectroscopy Laboratory, University of Rzesz{\'o}w,
Pigonia 1 Street, 35-310, Rzesz{\'o}w, Poland}

\begin{abstract}
We study, for the first time, the relations of two strong diffuse bands (DIBs) at 9633 and 9577~\AA, commonly attributed to C$_{60}^+$, to other interstellar
features seen in optical and UV spectra including H{\sc i}, Ca{\sc i}, Fe{\sc ii}, Na{\sc i}, Ti{\sc ii}, CN, CH, CH$^+$, and C$_2$ and DIBs 5780, 5797, 6196, 6269, 6284, and 6614.
 We analyzed 62 lines of sight where the stellar contamination by Mg{\sc ii} was corrected or found negligible for DIB 9633.
Equivalent widths of DIB 9577 were measured in 62 lines of sight. Poor mutual correlation between the strengths
of the above features and the major diffuse bands (5780 and 5797) as well as with other DIBs (with some exceptions) were revealed. The considered DIBs
are also poorly correlated with the features of neutral hydrogen, molecular carbon, and those of simple interstellar radicals. Perhaps this phenomenon can be explained if
the diffuse band 9577 is an unresolved blend of two or more interstellar features. There are indications that
9633 and 9577 diffuse bands  are stronger in $\sigma$-type clouds, i.e. these features resemble the
behavior of reasonably broad DIBs, which are strong in the lines of sight where the UV flux from the very hot nearby stars plays an important role.
\end{abstract}


\section{Introduction}

Two relatively strong and broad interstellar features near the diffuse bands (DIBs) 9577 and 9633 were first proposed to be possibly carried by the C$_{60}^+$
molecule by \cite{Foing1994}. The features are situated in the near-infrared (near-IR) region of the spectrum, heavily contaminated by telluric lines. A determination
of their exact profiles and strengths requires thus a telluric line divisor. Moreover, the DIB 9633 feature is inextricably intertwined with
the stellar Mg{\sc ii} line (\cite{Galazutdinov2000}; \cite{Galazutdinov2017}); the latter is negligibly weak only in the spectra of the hottest (O-type) stars.

\cite{Cami2010} reported the first detection of infrared emissions carried by neutral C$_{60}$, seemingly present on the surfaces of solid material (dust particles). It was observed only in the
vicinity of the peculiar planetary nebula Tc1. Independently,
\cite{Sellgren2010} reported the presence of neutral
C$_{60}$ in the NGC 7023 reflection nebula illuminated by the B star
HD\,200775.  The molecule was detected only  in the regions closest to the star.  All these
reports support the idea of a presence of the buckminsterfullerene cation in translucent interstellar clouds.

The reasonably recent publications by Campbell et al. (2015, 2016a, 2016b) restarted the discussion on whether the ``soccer ball'' molecule may carry the abovementioned near-IR spectral features. According to \cite{Campbell2015}, C$_{60}^{+}$  exhibits four relatively strong spectral
lines, centered at 9365.9$\pm$0.1, 9428.5$\pm$0.1, 9577.5$\pm$0.1, and 9632.7$\pm$0.1~\AA, with relative intensities of 0.2, 0.3, 1.0, and 0.8 respectively.

\cite{Galazutdinov2017} disputed the identifications, raising a problem
of the variable strength ratio of the two strong features. A precise determination of this ratio requires accurate elimination of telluric contaminations as well as that of the Mg{\sc ii} stellar line
from the DIB 9633 profile. The latter was done by means of calculating non-LTE synthetic spectra of the atmospheres of hot stars used to estimate the strength of
the contaminating line with respect to another non--blended one, near 4481~\AA\ (see details in Galazutdinov et al. 2017). The result was discouraging: the 9577/9633 strength ratio seems to be severely variable.

Another problem is the presence of weaker bands seen in laboratory spectra. Their relative equivalent width ratios should mimic the observed ones. Here
the problem of telluric contamination is even more severe.  The expected features are weak, while the telluric lines are strong; thus complete elimination of telluric contamination is not always
possible.  The identification of the C$_{60}^+$ molecule in translucent interstellar clouds was disputed by \cite{GK17a}; the observed strength ratios of the postulated C$_{60}^{+}$
features do not follow the laboratory predictions of \cite{Campbell2016}.  The observed strength ratios were
demonstrated to be variable, i.e. not the same from object to object. In their following publication, \cite{KM18} changed
the laboratory spectrum: the C$_{60}^+$ lines are at 9365.2$\pm$0.2, 9427.8$\pm$0.2, 9577.0$\pm$0.2, and 9632.1$\pm$0.2 with
relative intensities distributed as 0.26, 0.17, 1.00, and 0.84 respectively. Thus, the weak 9428 C$_{60}^+$ feature is weaker than that at 9365~\AA\ in contrast to the results reported in 2015.
This new spectrum matches the observed interstellar one better with one caution: the very weak 9428~\AA\ DIB is severely contaminated by telluric lines.

The latter problem should be solved if the observations are done outside the atmosphere. This is the subject of a recent publication by \cite{Cord19}. The authors used
the spectrograph onboard the Hubble Space Telescope (HST) to record the near-IR spectra of seven reddened and four unreddened stars. Despite their claim that C$_{60}^+$ is eventually discovered
beyond a doubt, the formerly missing feature (9428) is clearly seen only in BD +40 4220, albeit without detailed analysis of possible stellar contamination.
The presence of the 9428~\AA\ feature in other objects is doubtful. Moreover, the HST spectrograph
does not cover the second strongest C$_{60}^{+}$ DIB -- the 9633 one. Thus the authors of the latter paper were unable to analyze the relation between the two strong bands. We mentioned some problems with this in \cite{GK17a}. The lack of the major DIB 9633 in the observed wavelength range reduces the significance of the reported conclusions.

Apparently the statistics of the existing measurements are a serious problem. One needs more targets to convince everybody that the claimed (especially weak)
features really exist and that their strength ratios mimic the laboratory predictions. In this paper we present a sample of reddened targets
collected to check the possible relations of the proposed C$_{60}^+$ features to other interstellar ones. We emphasize that almost all interstellar features
are correlated, in many cases quite tightly \citep{MKHJ99}. Diffuse bands usually demonstrate good mutual correlations with
the 6196 -- 6614 relation being the ``champion'' \citep{KGBB16}. However, the latter publication warns that even a very tight correlation does not
guarantee that the two features share a carrier \citep{2020ApJ...899L...2K}.

\begin{figure}
    \includegraphics[width=12cm]{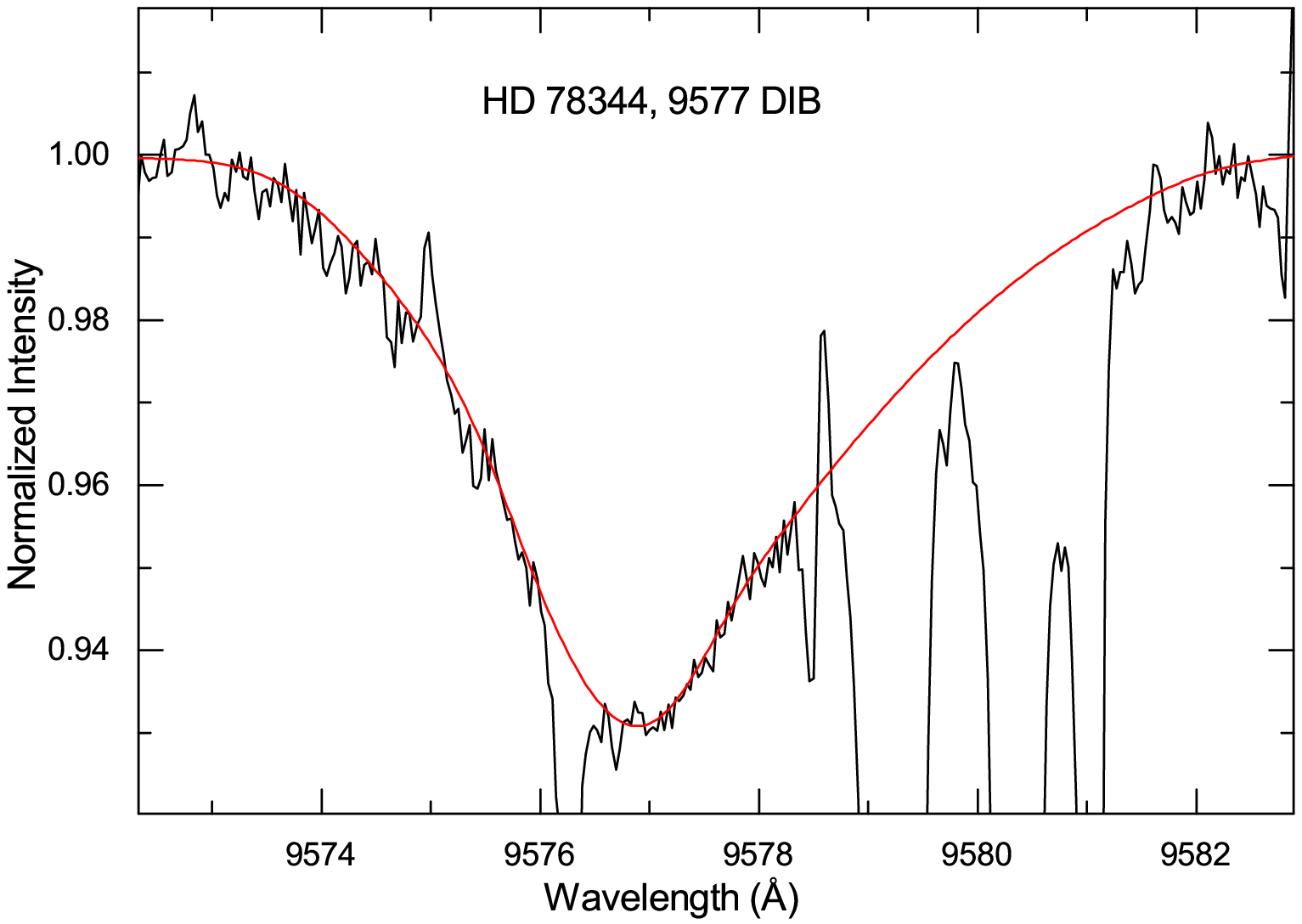}
    \caption{An example of the equivalent width measurement procedure with a manually set diffuse band profile (red).}
    \label{fit9577}
\end{figure}

\begin{figure}
    \includegraphics[width=12cm]{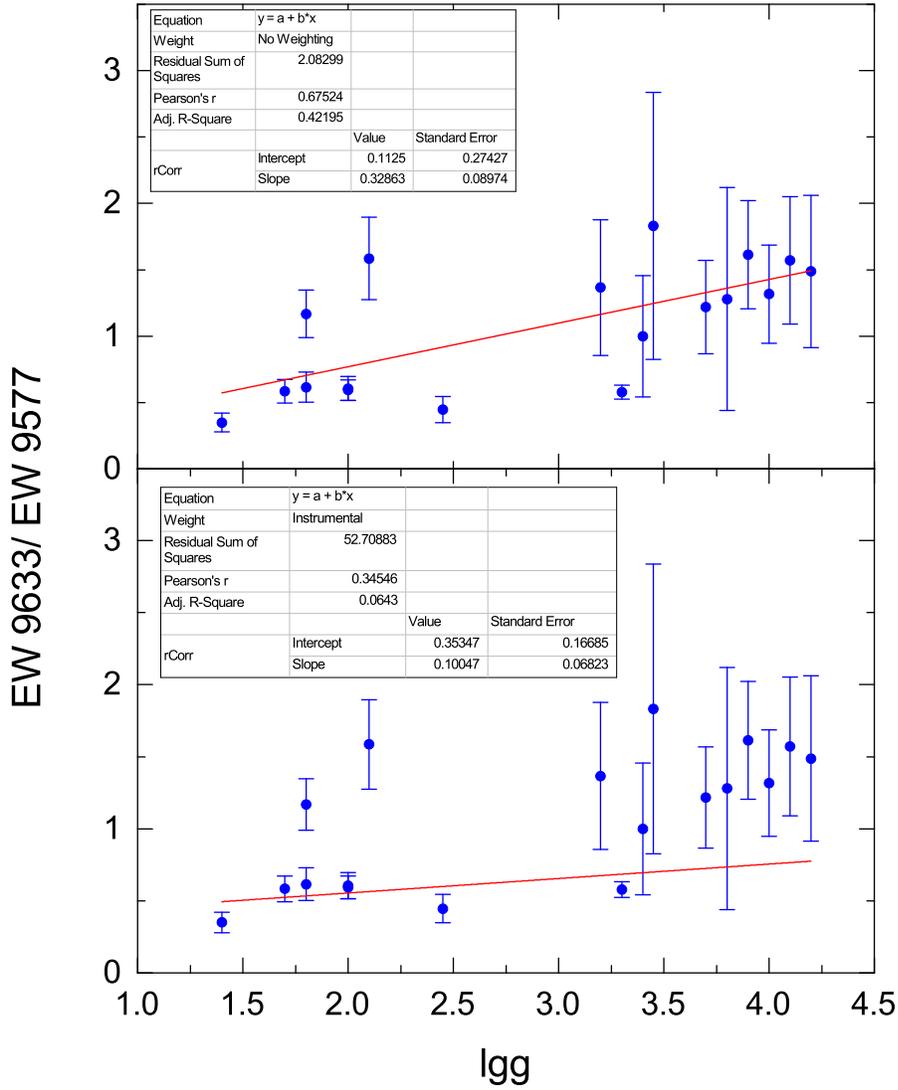}
    \caption{Fit over the {\it lg(g)} versus EW(9633)/EW(9577) relation with no error bars taken into account (top) and the weighted fit (bottom).
             Note the low correlation magnitude in the second plot.}
    \label{CorrNoErr}
\end{figure}

\begin{figure}
\center{
    \includegraphics[width=8cm]{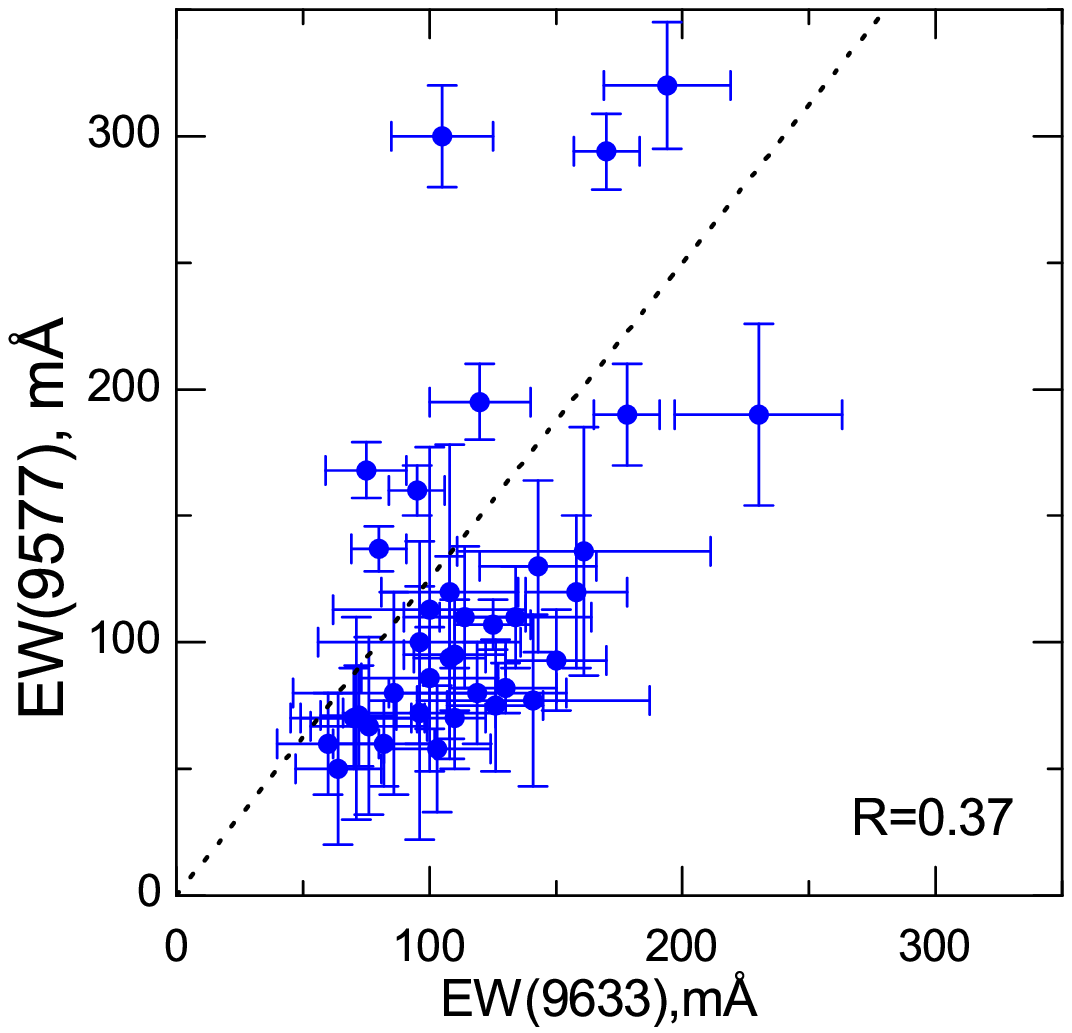}
    \vspace{2cm}
    \includegraphics[width=10cm]{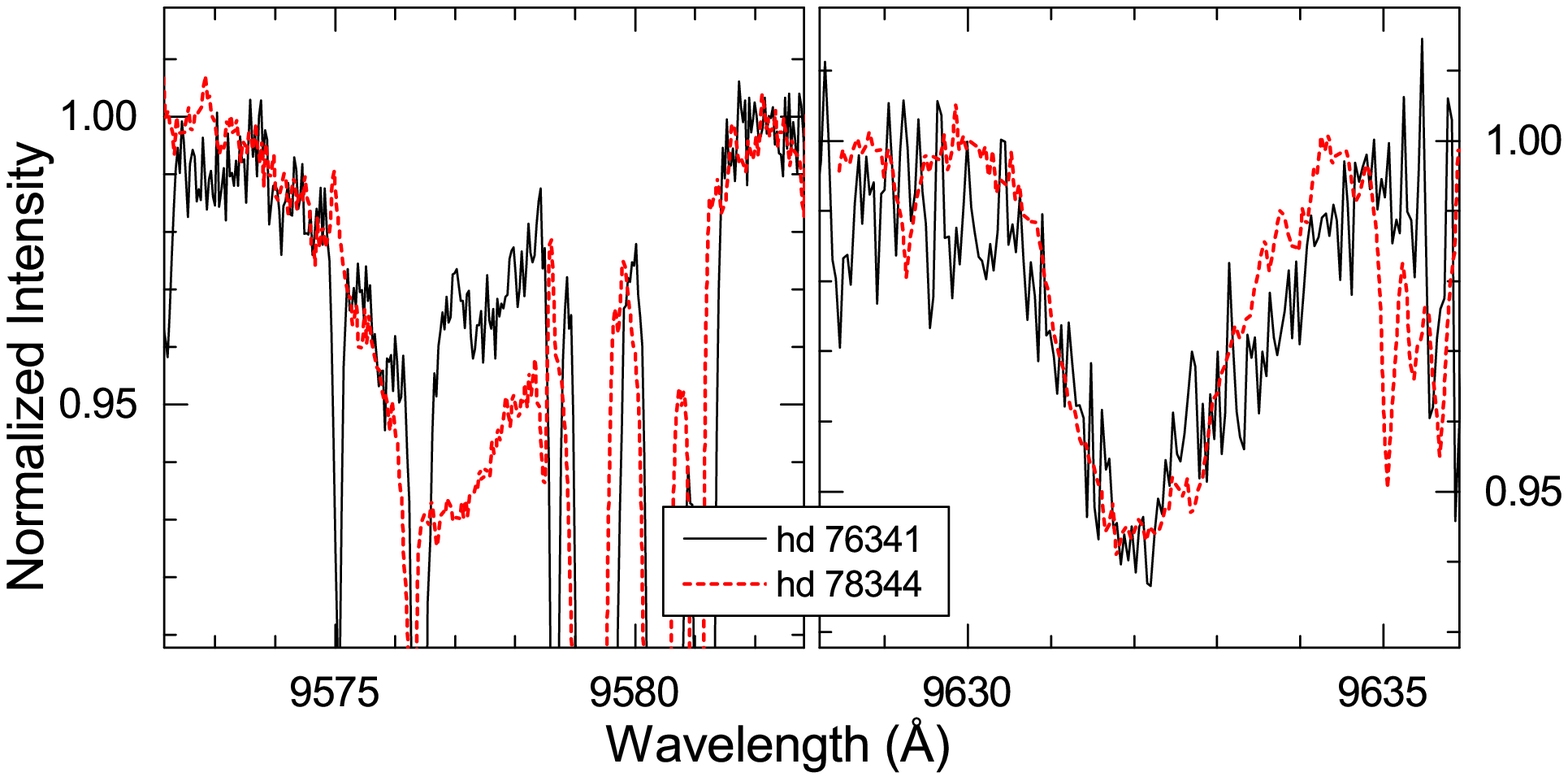}
    \caption{Top: correlation between the equivalent widths (m\AA) of the DIBs 9577 and 9633. Note the low correlation coefficient (R).
    The broken line represents the laboratory strength ratio of these bands.
    Bottom: an example of a different strength ratio for the DIBs 9633 and 9577 in two objects of similar spectral and luminosity class. }
    \label{mutual}
    }
\end{figure}

\cite{Walker2015} state that the weak
C$_{60}^+$ bands near 9366 and 9428~\AA\ can be traced in the spectra
of HD183143 (only 9366) and HD169454 (both). \cite{Walker2016} reported
the detection of diffuse bands 9632, 9577, 9428, 9365 and 9348
towards the stars HD 46711, HD 169454, HD 183143. In any case the conclusions were based on very small samples.
It is the natural consequence of the fact that IR DIBs (especially weak ones) can be traced only in the spectra
of heavily reddened targets.

In this paper we present an analysis of the largest sample so far of targets with
measured near-IR interstellar features attributed to C$_{60}^{+}$. We carefully corrected not only the telluric line contamination but also that caused by stellar
lines (Galazutdinov et al. 2017). The latter contamination poses special difficulties while dealing with broad
interstellar features with a width comparable to stellar lines.  Here we use the results from Galazutdinov et al. (2017) supplemented with measurements
of hot objects lacking a strong Mg{\sc ii} effect.

We also analyze possible relations between the two strong infrared bands (C$_{60}^{+}$) and the other well-known DIBs, as well as abundances of simple interstellar molecules and atoms.

\section{Spectral data}

Our observations have been collected using several high
resolving power echelle spectrographs. They are as follows.
\begin{itemize}

\item The Ultraviolet and Visual Echelle Spectrograph (UVES) is fed with the 8m Kueyen VLT mirror (ESO, Paranal, Chile; \citep{Dekal00}). The
    spectral resolving power reaches R=80,000 in the blue range and R=110,000
    in the red. The telescope size allows one to get high signal-to-noise ratio (S/N) spectra even for rather faint stars.
\item The Fiber-fed Extended Range Optical Spectrograph (FEROS) is fed with the 2.2m MPG/ESO telescope (ESO, LaSilla) \citep{Kauf99}. It allows one to record in a single exposure a spectral range of 3600 to
    9200~\AA\, divided into 39 echelle orders. The resolving power of Feros
    spectra is R=48,000. Feros spectra cover a broad wavelength range per each order,
    which makes the spectrograph a very useful tool for
    checking the spectral types and luminosity classes of the observed
    targets and for measuring DIBs and atomic/molecular features. However, the C$_{60}^{+}$ features are out of its range.
\item The Echelle SpectroPolarimetric Device for the Observation of Stars (ESPaDOnS) spectrograph\footnote{For details see https://www.cfht.hawaii.edu/Instruments/Spe\-ctroscopy/Espadons/} is a bench-mounted high resolving power echelle     spectrograph/spec\-tro\-pola\-ri\-meter, attached to the 3.58m   Canada-France-Hawaii telescope at Maunakea (Hawaii, USA). It is
    designed to obtain a complete optical spectrum in the range of
    3700 to 10050~~\AA.
    The whole spectrum is divided into 40 echelle
    orders. The resolving power is about 68,000. Spectra from
    ESPaDOnS were obtained during the runs 05Ao5 (in 2010, PI B.~Foing) and 15AD83 (in 2015,  PI G.~Walker).
    The high altitude (more than 4 km) and exceptionally low humidity on the site make Mauna Kea an ideal
    place for the study of IR DIBs. The quality of telluric line removal in ESPaDOnS spectra is the best among
    all the spectral data we have used. Unfortunately, the major DIB 9633 is out of the observed spectral range in the spectra from ESPaDOnS.
    This explains why the number of the DIB 9633 measurements is lower than that for 9577 (Tab. 1).
\item
    The Bohyunsan Echelle Spectrograph (BOES) of the Korean National Observatory \citep{kimetal2007} is installed at the 1.8m telescope of the Bohyunsan Observatory in Korea. The spectrograph has three
    observational modes allowing resolving powers of 30,000, 45,000, and
    90,000. In any mode, the spectrograph covers the whole spectral range of
    $\sim$3500 to $\sim$10,000~\AA, divided into 75 -- 76 spectral
    orders.
\item
    The MAtrix Echelle SpecTROgraph  (MAESTRO; \citep{1999KPCB...15..216M}) is attached to the 2m
    telescope at the Terskol Observatory (the North Caucasus, Russia). It is a
    three branch cross-dispersed echelle spectrograph installed at the Coud{\'e}
    focus (F/36) of the telescope. It was designed for stellar spectroscopy
    using high resolutions ranging from R = 45,000 to 190,000 in the 3500 -- 10000 \AA\ spectral region. The lowest resolving power mode (sufficient for our     programme) allows one to obtain spectra of targets as faint as $\sim$10$^m$ with a sufficient ($\sim$100) S/N;
\end{itemize}

All spectra were processed and measured in a standard way using both IRAF \citep{Tody1986} and our own DECH\footnote{http://www.gazinur.com/Download.html} codes.
The spectral resolutions provided by the abovementioned instruments are not identical but all are high enough to precisely measure the strengths of
atomic and molecular interstellar lines and, especially, of the broad DIBs.

Measuring the DIBs 9633 and 9577 is not a simple procedure because of strong telluric line remnants and, sometimes, a high level of noise.
In most cases, to measure the equivalent width, we used the manual profile fit method: a cubic spline fit over manually set profile points
(see Fig. \ref{fit9577}). %  \sout{in the electronic version of the article}
The method provides the possibility to measure lines of any complex, irregular shape. The resulting data for the DIBs 9577 and 9633  are presented in Tab.1.
The equivalent width errors were estimated using Eq. 7 from Vollmann \& Eversberg (2006) where both spectral noise and
uncertainty of the continuum normalization are taken into account.

\subsection{Contamination of the DIBs 9633 and 9577 by telluric and stellar spectra}

The near-IR wavelength range suffers strong contamination by telluric lines. To eliminate the contamination we applied the
classical method based on the use of a divisor --- a spectrum of a hot, unreddened, and preferably rapidly rotating star.
The method permits one to adjust both the positional and intensity differences of the telluric lines in the studied object and divisor
spectra. We used several stars as telluric line divisors, always trying to observe the one closest to the chosen target. As a divisor we used
spectra of Spica (HD\,116658, B1V), HD120315 (B3V), HD218045 (B9III) and some other targets with characteristics
satisfying the requirements given above. Our method is described in detail by \cite{Galazutdinov2017}. Here we used the same procedures, only the number of
analyzed targets is more than twice as large as in \cite{Galazutdinov2017}.

Complex and variable spectra of the chosen divisors, when used for the removal of telluric lines, may introduce unwanted distortions into the
resulting profiles of relatively broad bands, e.g. 9633, since real spectra contain stellar lines. On the other hand the
telluric lines (mostly of the atmospheric H$_2$O) are in many cases saturated, which makes their removal practically impossible.
Thus some remnants of the telluric spectrum are still seen in our resultant spectra. With such
small contamination we can still reliably measure the chosen DIBs (manually fitting the profile to the undisturbed points).
However, in many cases the telluric contamination is the main source of uncertainty for the equivalent width measurements given in Tab.~1.

It is worth mentioning that the stellar contamination of the DIB 9633 depends on the effective temperature
being relatively large in late-B type atmospheres (like HD183143) and small in very
hot O stars (like HD76341). The problems of the elimination of the stellar Mg{\sc ii}
line from the profile of the DIB 9633  was discussed broadly in \cite{Galazutdinov2017}.
 The applied method was criticized by \cite{Lall18} who claimed that the ratio of
the two strong features,  9577/9633 correlates with the surface
gravity of the star (see their Fig. 1), i.e. the stellar contamination effect is not properly addressed.
However, the linear fit, shown in the abovementioned figure, was calculated neglecting the weighting procedure by the individual error bars. If the straight line is calculated using proper weights the trend disappears,
i.e. the determinations of stellar parameters by \cite{Galazutdinov2017} are correct and we use these values also in the current paper  (see Fig. \ref{CorrNoErr}).
However, to improve the statistics, we have added  additional targets. Some of them are very hot stars (similar to HD76341) where the Mg{\sc ii} contamination to the
DIB 9633 profile is negligibly small, although the equivalent width of the DIB 9633 may be slightly overestimated.
For cooler targets we have measured the DIB 9577 only.  It is of basic importance to check whether the strength ratio
of the two strong  DIBs 9577 and 9633 is close or not to the laboratory
predictions. If not, the idea that C$_{60}^{+}$ is their carrier must be postponed until we have a lab spectra of C$_{60}^{+}$ with a variable ratio of these major features due to the variation of some physical/chemical parameter, e.g. rotational temperature.

\begin{table*}
\caption{Observed targets and equivalent widths of the DIBs 9633 and 9577 (m\AA).  Spectral type, luminosity class and \textit{v}sin\textit{i}
are taken from the SIMBAD database. T$_{eff}$, {\it log g} are from Galazutdinov et al. (2017).}
\label{correlationtab}
\tiny
\center{
\begin{tabular}{lccrrc|rccrr}
\hline
\hline
Object      & SpL or Teff/lgg & \textit{v}sin\textit{i}&9633& 9577
                                                           &&     Object & SpL or Teff/lgg & \textit{v}sin\textit{i}&9633    & 9577 \\
\hline
BD-14 5037  &  18000/1.8   & 42 &125$\pm$15 &  107$\pm$10  &&    HD148379    &  17000/1.7   & 51 & 80$\pm$11 &  137$\pm$9  \\
CD-32 4348  &  19500/2.45  & 36 & 75$\pm$16 &  168$\pm$11  &&    HD148605    &  20500/4.2   & 145&119$\pm$35 &   80$\pm$20  \\
BD+40 4220  & O6.5f+O5.5f  &$>$200&263$\pm$62& 350$\pm$78  &&    HD148937    &  O6fp        &    &143$\pm$23 &  130$\pm$34  \\
BD+59 2735  &  B0Ib        &    &163$\pm$14 &  368$\pm$150 &&    HD149038    &  O9.7Iab     & 52 &100$\pm$27 &   86$\pm$20  \\
Cyg OB2 7   &  O3If        & 75 &163$\pm$14 &  235$\pm$55  &&    HD149757    &  O9.2IVnn    & 303& 96$\pm$30 &   72$\pm$50  \\
Cyg OB2 8   &  O6Ib+O4.5III&    &           &  200$\pm$45  &&    HD150136    &  O4III+O8    &    &108$\pm$27 &  120$\pm$58  \\
Cyg OB2 12  &  B3Iae       &    &           &  390$\pm$70  &&    HD151804    &  O8Iaf       & 72 &126$\pm$19 &   75$\pm$26  \\
HD13256     &  B1Ia        &    &           &   95$\pm$54  &&    HD152408    &  O8Iape      &    &110$\pm$20 &   95$\pm$22  \\
HD22951     &  B0.5V       & 10 & 72$\pm$22 &              &&    HD152424    &  OC9.2Ia     & 59 &161$\pm$50 &  136$\pm$49  \\
HD23180     &  24000/3.45  & 78 &141$\pm$46 &   77$\pm$34  &&    HD153919    &  O6Iafcp     &    &114$\pm$24 &  110$\pm$28  \\
HD27778     &  15500/3.8   & 92 & 64$\pm$17 &   50$\pm$30  &&    HD155806    &  O7.5V       & 52 & 86$\pm$40 &   80$\pm$40  \\
HD36861     &  O8III       & 52 & 71$\pm$22 &   70$\pm$40  &&    HD167264    &  29000/3.2   & 82 & 82$\pm$20 &   60$\pm$17  \\
HD37022     &  O7Vp        & 29 & 72$\pm$15 &   71$\pm$20  &&    HD167971    &O8Iaf(n)+O4/5 & 65 &178$\pm$13 &  190$\pm$20  \\
HD37041     &  O9.5IVp     &134 &108$\pm$14 &   94$\pm$40  &&    HD168607    &  B9Iaep      &    &           &  335$\pm$120 \\
HD40111     &  B0III       &141 & 76$\pm$23 &   67$\pm$35  &&    HD168625    &  B6Iap       &    &194$\pm$25 &  320$\pm$25  \\
HD54662     &  O7Vzvar     & 95 & 60$\pm$20 &   60$\pm$20  &&    HD169454    &  21000/2.1   & 39 &130$\pm$20 &   82$\pm$10  \\
HD55879     &  O9.7III     & 26 & 57$\pm$20 &              &&    HD170740    &  21000/3.9   & 40 &150$\pm$20 &   93$\pm$20  \\
HD57061     &  O9II        & 51 & 65$\pm$17 &   94$\pm$47  &&    HD183143    &  11500/1.4   & 37 &105$\pm$20 &  300$\pm$20  \\
HD76341     &  34000/3.7   & 66 &134$\pm$30 &  110$\pm$20  &&    HD184915    &  27000/3.4   & 220& 70$\pm$25 &   70$\pm$20  \\
HD78344     &  31000/3.3   & 98 &170$\pm$13 &  294$\pm$15  &&    HD190603    &  B4 Ia       & 45 &155$\pm$111&  150$\pm$20  \\
HD80077     &  17000/2.0   & 47 & 95$\pm$11 &  160$\pm$10  &&    HD194279    &  B5Ia        & 48 &           &  161$\pm$52  \\
HD91824     &  O7V         & 51 & 60$\pm$33 &   84$\pm$28  &&    HD204827    &  O9.5IV      &    &           &   10$\pm$10  \\
HD104705    &  B0III/IV    &    & 96$\pm$40 &  100$\pm$40  &&    HD208501    &  B8Iab       & 41 &           &   97$\pm$54  \\
HD113904    &WC5+B0III+O9IV& 106&100$\pm$38 &  113$\pm$64  &&    HD219287    &  B0Ia+       &    &           &  144$\pm$74  \\
HD136239    &  17000/1.8   & 43 &120$\pm$20 &  195$\pm$15  &&    HD226868    &  O9.7Ib      &106 &           &  129$\pm$38  \\
HD143275    &  B0.3IV      &165 & 94$\pm$26 &              &&    HD228712    &  B0.5Ia      &    &           &  222$\pm$52  \\
HD144470    &  B1V         & 95 &103$\pm$21 &   58$\pm$25  &&    HD228779    &  O9Ia        &    &           &  104$\pm$32  \\
HD145502    &  21000/4.0   & 98 &158$\pm$20 &  120$\pm$30  &&    HD229059    &  B2Iab       &    &           &  205$\pm$35  \\
HD147165    &  B1 III      & 22 &230$\pm$33 &  190$\pm$36  &&    HD235825    &  O9IV        &    &           &  110$\pm$40  \\
HD147888    &  16000/4.1   & 104&110$\pm$12 &   70$\pm$20  &&    HD254577    &  B0.5Ib      &    &           &  185$\pm$90  \\
HD147889    &  B1.5V       & 100&           &   26$\pm$18  &&    HD281159    &  binary      &584 &           &  227$\pm$82  \\
\hline
\end{tabular}
}
\end{table*}

\begin{table}
\caption{Correlation coefficient R estimated for {\bf n} pairs}
\label{correlationR}
\begin{tabular}{|lccc|lcc|}
\hline
\hline
pair of DIBs          &  R    & n   &  &     pair  of DIBs         &  R    & n   \\
\hline
9633/9577      &  0.37 & 37  &  &     9633/NaI 3303  &  0.23 & 36  \\
9633/E(B-V)    &  0.50 & 45  &  &     9577/NaI 3303  &  0.01 & 33   \\
9577/E(B-V)    &  0.47 & 60  &  &     9633/TiII 3242 &  -0.25 & 32   \\
9633/6196      &  0.54 & 45  &  &     9577/TiII 3242 &  0.02 & 30   \\
9577/6196      &  0.70 & 62  &  &     9633/FeI 3860  &  0.02 & 41   \\
9633/6284      &  0.25 & 45  &  &     9577/FeI 3860  &  0.30 & 38   \\
9577/6284      &  0.65 & 62  &  &     9633/CN 3874.6 &  0.40 & 41   \\
9633/5780      &  0.44 & 44  &  &     9577/CN 3874.6 &  0.11 & 39   \\
9577/5780      &  0.71 & 61  &  &     9633/CaI 4227  &  0.09 & 43   \\
9633/5797      &  0.36 & 43  &  &     9577/CaI 4227  &  0.27 & 43   \\
9577/5797      &  0.52 & 60  &  &     9633/CH$^+$ 4232  &  0.44 & 44  \\
9633/6614      &  0.37 & 44  &  &     9577/CH$^+$ 4232  &  0.36 & 57  \\
9577/6614      &  0.77 & 61  &  &     9633/CH 4300   &  0.25 & 44  \\
9633/6269      &  0.32 & 45  &  &     9577/CH 4300   &  0.22 & 57  \\
9577/6269      &  0.66 & 62  &  &                    &       &      \\
\hline
\end{tabular}
\end{table}

\section{Results}

Using our sample of reddened stars we measured the DIB 9577 in 62 objects and the DIB 9633 in 43 objects. In some of our targets the latter DIB is not available or
is contaminated by the Mg{\sc ii} stellar line \citep{Galazutdinov2017}. In our sample of DIB 9633 measurements we included only the hottest stars where the stellar contamination is negligible. If both the 9633 and 9577 bands, attributed
to C$_{60}^{+}$, share the same carrier, their ratio should be similar to that obtained in the laboratory measurements and thus
their strengths should correlate tightly. The correlation
between the two features, shown in Fig. \ref{mutual}, is surprisingly poor in contrast to the expectations. Moreover, the correlation coefficient between the DIBs 9633 and 9577 is
among the lowest for any pair of diffuse bands.
Perhaps this can be partially explained
by the relatively high uncertainty of the measured equivalent width due to the imperfection of the telluric line removal procedure.
However, as is shown in the bottom part of Fig.\ref{mutual}, the variability of the 9577/9633 ratio cannot be completely explained by the influence of telluric lines.
It is worth mentioning that all of the correlation coefficients were calculated with the measurement error $\sigma$ taken as a weight in the form of 1/$\sigma^2$.

The poor correlation between the DIBs 9577 and 9633  casts doubts on whether they are of the same origin, i.e. whether they do represent
the spectrum of the C$_{60}^{+}$ molecule. We now check to see how both DIB carriers are possibly related to other interstellar species.

\begin{figure}
    \includegraphics[width=12cm]{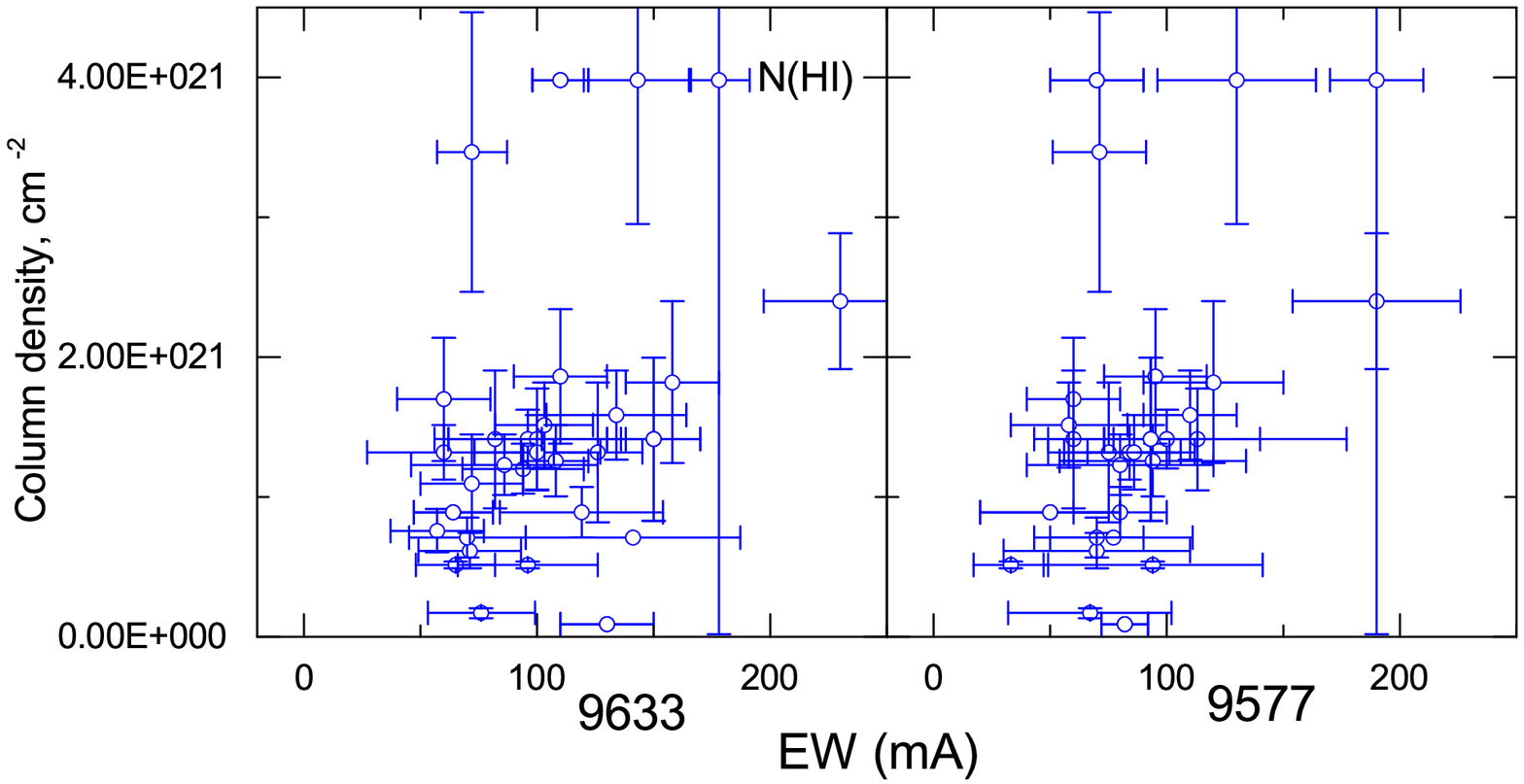}
    \caption{The lack of correlation between the equivalent widths of DIBs 9633, 9577,  and the column density of neutral hydrogen.(The complete figure set (5 images) is available in the online Journal.)}
    \label{H1}
\end{figure}

\begin{figure}
    \includegraphics[width=12cm]{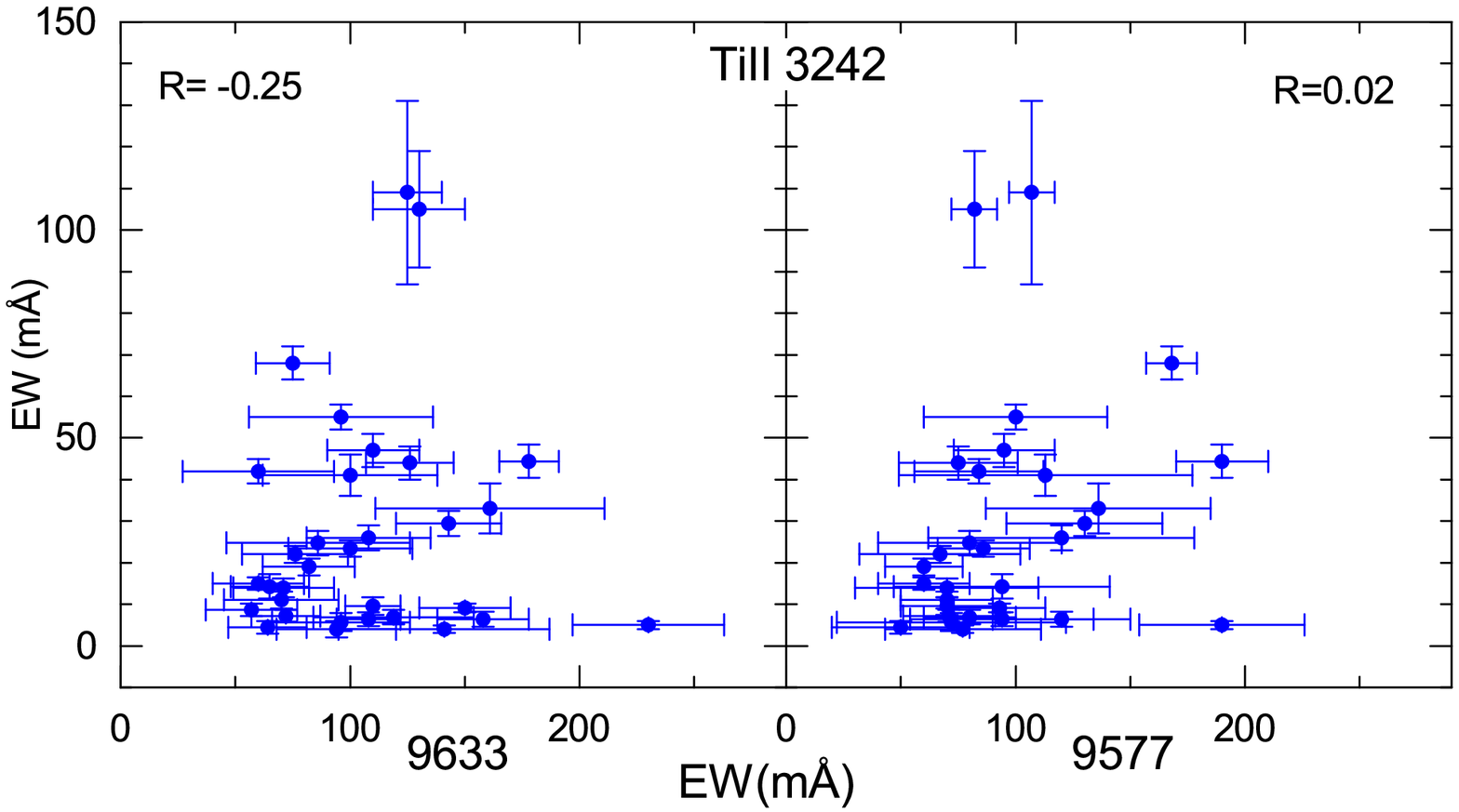}
    \caption{The same as Fig. \ref{H1} but for Ti{\sc ii}. R is the correlation coefficient.}
    \label{Ti2}
\end{figure}

It is certainly of interest to investigate whether C$_{60}^{+}$ is mixed well with the interstellar atomic hydrogen. In Fig. \ref{H1} we try to
relate both DIBs to the hydrogen column density taken from the compilation  of Gudennavar et al.(2012), based on measurements of absorptions observed by space-born instruments.
As seen in Fig. \ref{H1} both of the considered DIBs hardly correlate with the hydrogen column density.  The correlation coefficients are 0.35 and 0.46 for the DIBs 9633 and 9577, respectively.
Apparently the DIB carriers are not evenly distributed in the interstellar H{\sc i} clouds and the H{\sc i} column density does not allow one to predict the intensities of both DIBs. The same situation
is observed in the cases of all atomic gases: Ti{\sc ii}, Na{\sc i}, Ca{\sc i} or Fe{\sc i}; for all these relations the correlation coefficients are very small (see the online figure set associated with Fig. \ref{H1}).

\begin{figure*}
    \includegraphics[width=16cm]{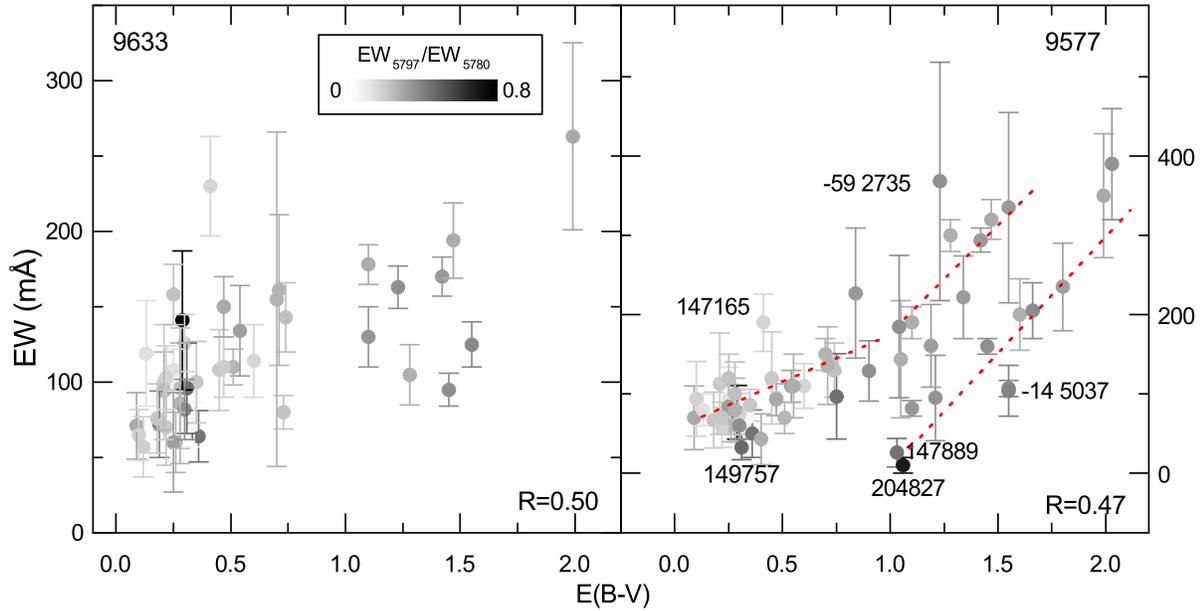}
    \caption{The correlation between the equivalent widths (in m\AA) of the DIBs 9577, 9633 and E(B-V). Note the low correlation coefficients (R).
     The relation for 9577 may suggest two sequences, each thus correlating more tightly (see the text). }
    \label{ebv}
\end{figure*}

Another interesting question is whether our DIBs are related in some way to interstellar dust. The optical depth of the latter is usually measured using the color
excess E(B-V). Traditionally a feature that correlates with E(B-V) is considered as interstellar.  Fig. \ref{ebv} relates both strong DIBs to E(B-V) and
proves a rather poor correlation between the DIB carriers and the dust grains. The color intensity of the symbols in the plots is associated with the equivalent width ratio of the
major DIBs at 5797 and 5780 \AA, i.e. darker colors correspond to a stronger $\zeta$ effect, i.e. a higher equivalent width (5797/5780)) ratio.
It is evident that $\zeta$-type objects have a tendency to present weak  C$_{60}^{+}$ associated DIBs.
A rather speculative issue is the possible presence of two sequences in the relation between the equivalent width of the DIB 9577 and the reddening, roughly marked by the red dashed lines
in the right panel. This possible split cannot be explained by $\zeta$/$\sigma$ progressions. Indeed, a heavily reddened $\zeta$-type object HD204827 and the progenitor of
$\zeta$-type clouds HD149757 both exhibit a very weak DIB 9633. If any, the split can be explained, for example, by a presence of two independent diffuse bands at $\lambda$9633.
However, a much more representative sample of measured objects is necessary to confirm or reject this hypothesis.

\begin{figure}
    \includegraphics[width=12cm]{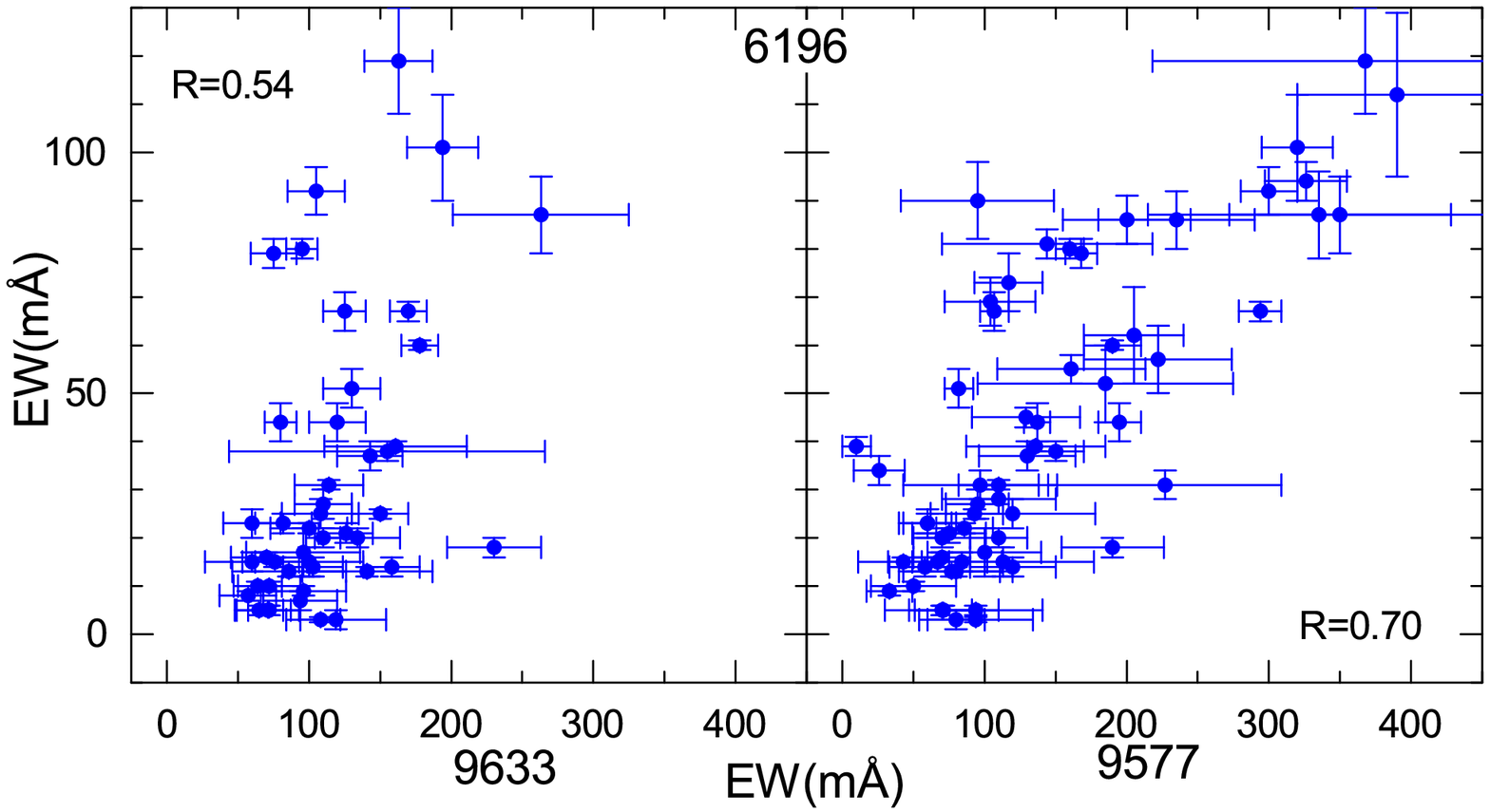}
    \caption{The correlation between the equivalent widths (in m\AA) of the DIBs 9577, 9633 and the equivalent width of the narrow DIB 6196 (in m\AA). The correlation coefficient for 9577 (R=0.70) is among the highest in our sample.
    \textbf{The complete figure set (6 images) is available in the online Journal.}}
    \label{6196}
\end{figure}

As mentioned in the Introduction, diffuse bands usually do correlate quite tightly. Even
the major DIBs  5780 and 5797~\AA, well known as being of a very variable depth ratio (\citealt{KW88}; \citealt{2019MNRAS.486.3537K}),
exhibit reasonably well a correlation of equivalent widths if a sufficiently large sample is considered (\citealt{2016A&A...585A..12B}).
How are the DIBs 9577 and 9633 related to the other ones? Fig. \ref{6196} depicts their relations to the narrow DIB 6196. The latter is known
to correlate very tightly with practically all optical DIBs, especially with the one at 6614. Both near-IR DIBs, but 9577 especially, do correlate reasonably tightly with the narrow optical DIBs. However, the correlation is poor in comparison with those between all possible optical DIBs \citep{MKHJ99}. Other DIBs usually correlate less tightly with the suspected C$_{60}^{+}$ DIBs.
9577 is also quite well correlated with the DIB 5780 (R=0.71). However, the same correlation coefficient for 9633 is only 0.44. All correlation coefficients
are listed in Tab. 2. Generally they are smaller for the DIB 9633 than for DIB 9577, although the DIB 9633 sample size is systematically smaller.  This is another argument
for the binary origin of the DIB 9577: like the cumulative effect of the average of many clouds increases the magnitude of mutual correlation between diffuse bands, the blending
of diffuse bands smears the peculiarities and, might provoke an increasing correlation coefficient.

\begin{figure}
    \includegraphics[width=12cm]{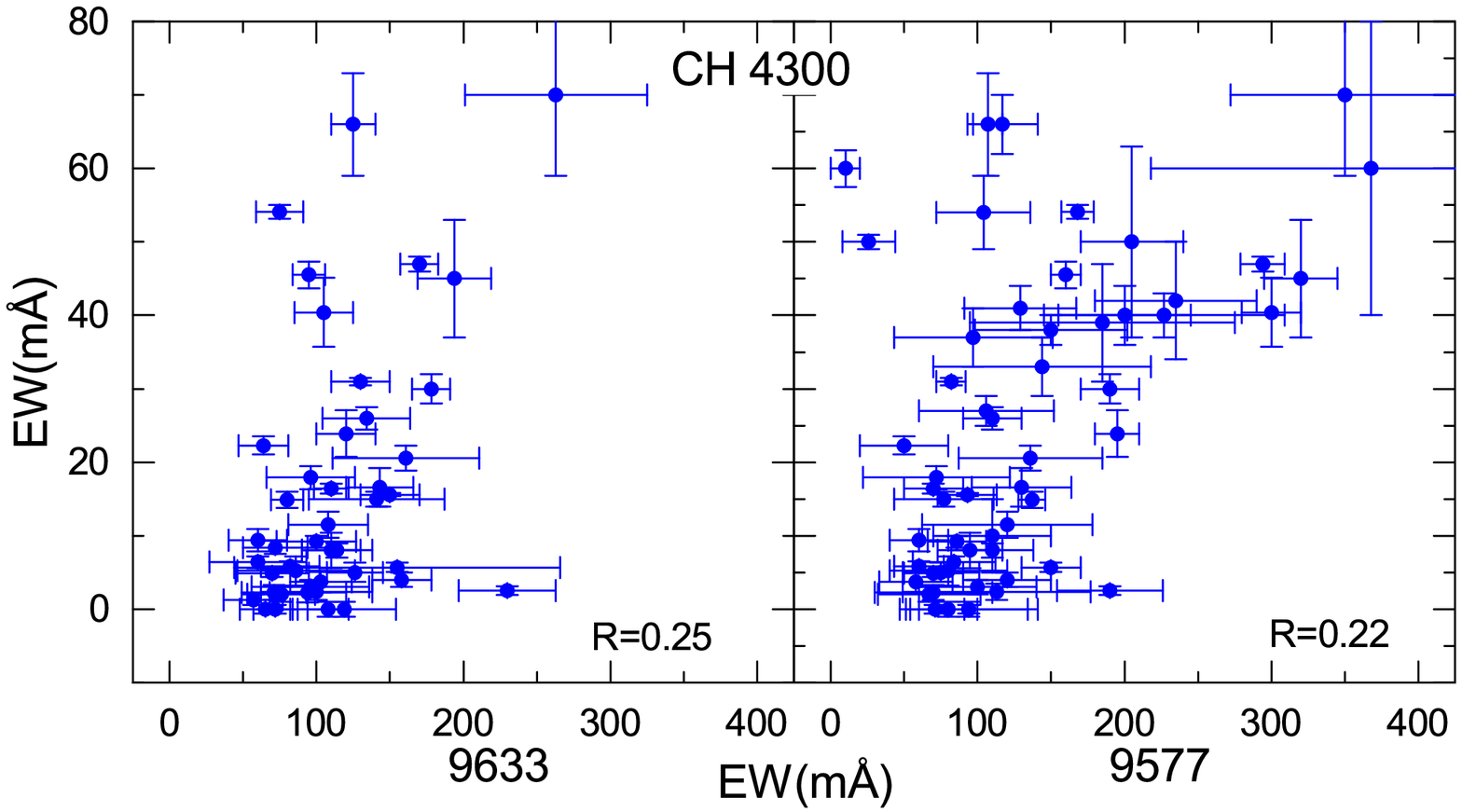}
    \caption{The correlation between the equivalent widths (in m\AA) of the DIBs 9577, 9633  and the equivalent width of the CH 4300~\AA\ line. R is the correlation coefficient.
    \textbf{The complete figure set (3 images) is available in the online Journal.}}
    \label{CH4300}
\end{figure}

\begin{figure}
    \includegraphics[width=12cm]{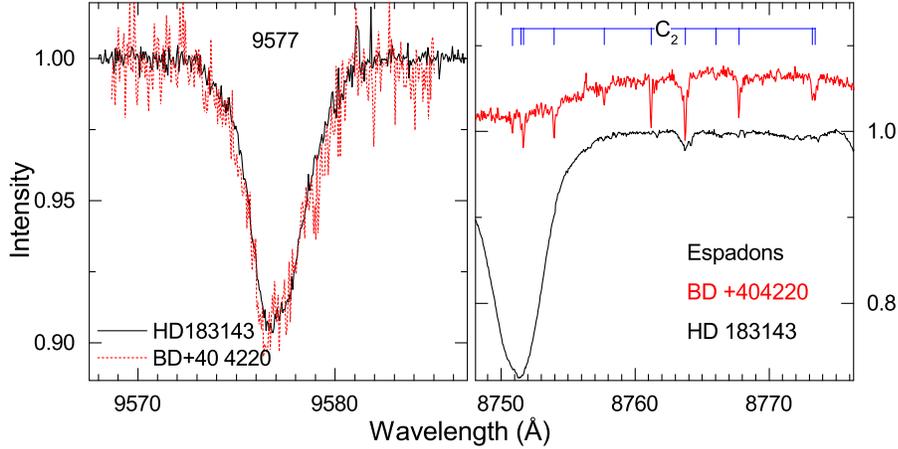}
    \caption{An example of poor correlation of the DIB 9577  with molecular carbon. The profile of 9577 is identical in the spectra of two heavily reddened stars while the intensity of the C$_2$ bands differs drastically.
    Also, this figure is a good example of a comparison of the DIB 9577  profiles observed in spectra of fast (BD+40 4220) and slow (HD 183143) rotating stars.
    }
    \label{c2}
\end{figure}

It is of interest whether the considered DIBs show any correlation with interstellar features of simple radicals. Fig. \ref{CH4300} illustrates
the relation between the two DIBs and the strongest line of a simple CH molecule. The correlation coefficient is 0.25 and 0.22 for the DIBs 9633 and 9577,  respectively, which is in fact
negligible. Apparently the DIB carriers are not well mixed with the simple radicals. The correlations for CH$^+$ and CN are not much better.
 On other hand, the correlation of DIBs 9577,9633 with CH$^+$ is almost two times higher than that with CH. Perhaps this is not just a statistical fluctuation but an
observational fact supporting our idea that these DIBs are stronger in $\sigma$-type clouds.
The C$_2$ radical seems to be the most interesting, as such small, carbon molecules may be considered as building blocks for more complex species such as C$_{60}^{+}$. Fig. \ref{c2} shows that
apparently the abundance of C$_2$ is not related in any way to C$_{60}^{+}$. This casts additional doubt on the identification of the  DIBs 9577 and 9633 as carried by C$_{60}^{+}$.

\begin{figure}
    \includegraphics[width=12cm]{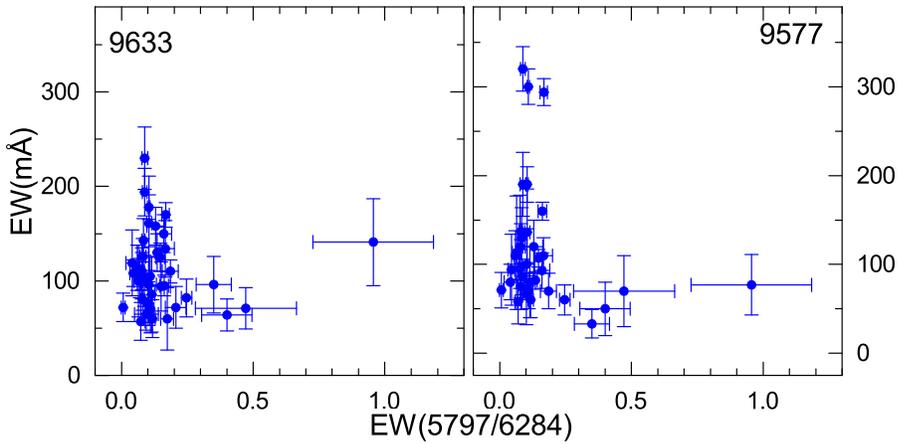}
    \caption{The plots divide our targets into two groups following either the strong narrow DIBs (5797) or the broad ones (6284). }
    \label{zeta-effect}
\end{figure}

Since 1988 \citep{KW88} interstellar clouds are being divided into $\sigma$ and $\zeta$ types. This division is based on the strength ratio
of the major DIBs: 5780 or 6284 and 5797 or 6379. It is thus interesting how the intensities of 9577 and 9633 react to the 5797/6284 ratio. This is illustrated in Fig. \ref{zeta-effect}.
The points in this figure are distributed in the form of the letter "L". Narrow diffuse bands may be very strong when our so-called C$_{60}^{+}$ DIBs
are very weak. The opposite situation is possible as well. However, both sequences intercept each other.
 This may suggest that DIBs 9577 and 9633 are stronger in $\sigma$-type clouds, i.e. the carriers of these features follow the
intensity of reasonably broad DIBs which are strong in the lines of sight where the UV flux from the very hot nearby stars plays an important role.

\section{Conclusions}

Our considerations allow us to infer the following conclusions.
\begin{itemize}
\item
The two strong, so-called C$_{60}^{+}$ DIBs (9577 and 9633), exhibit a poor mutual correlation, which casts doubt on their common origin.
However the lack of correlation can be explained, e.g. if the DIB 9577 is a blend of two features (see Fig. \ref{ebv} and the text).
\item
 The relation of the two DIBs to E(B-V) is also quite poor.  Again, this might be due to a possible blending effect in DIB 9577.
\item
The intensities of the two strong DIBs are poorly related to the H{\sc i} column density and, generally, to all molecular/atomic features analyzed in this article,
although the correlation coefficients vary from feature to feature (see Tab. 2). In particular, the two DIB carriers do not seem to be related to simple interstellar
radicals such as CH but also CH$^+$ and CN. A particular difficulty concerns C$_2$ as the latter may be considered as a building block for C$_{60}$.
\item
Also, the DIBs 9633 and 9577  are stronger in $\sigma$-type clouds, i.e. the carriers of these features follow the
intensity of reasonably broad DIBs, which are strong in the lines of sight where the UV flux from very hot nearby stars plays an important role.
An additional argument in favor of this suggestion is the correlation of DIBs 9577,9633 with CH$^+$, quite weak though but twice as high as that with CH.
\end{itemize}

Poor mutual correlation of the 9577 and 9633 diffuse bands  as well poor correlations of these DIBs with other interstellar features may be partially explained by the
influence of: (i) imperfection of the telluric line removal procedure, (ii) contamination by stellar lines, and finally, by (iii) an unresolved blend with other DIBs.
The remedy for the first two issues is evident though not easy to perform: extensive observation of both reddened and unreddened targets with the aid of
space telescopes, with subsequent correction of the intrinsic parameters of the stellar lines in order to precisely model the stellar spectra.
The third issue might be quite difficult to resolve without essentially increasing measurement quality and the quantity of the studied lines of sight.
Indeed, it is difficult to detect and measure precisely the moderately reddened targets in the near-IR because of the weakness
of near-IR DIBs in objects lacking an essential amount of dust.

There are widely spread reports in favor of the identification of interstellar  C$_{60}^{+}$ (see, e.g. Linnartz et al. 2020, Woods 2020). However, in our opinion the identification cannot be decisively recognized without explaining the issues raised from the observational facts presented in this article.

\begin{acknowledgements}
This paper includes data gathered with the VLT and UVES spectrograph, programs 067.C-0281(A), 082.C-0566(A), 092.C-0019(A).
The authors are grateful to Dr. Jacco Th. van Loon for the careful reading of the paper and the valuable suggestions and comments.
G.A.G., G.V. and N.R.I. acknowledge the support of Ministry of Science and Higher Education of the Russian Federation under the grant 075-15-2020-780 (N13.1902.21.0039).
J.K. acknowledges the financial support of the Polish National Science Center \textemdash the grant UMO-2017/25/B/ST9/01524 for the period 2018 \textendash\ 2021.
G.A.G. and J.K. acknowledge the Chilean fund CONICYT grant REDES180136 for financial support of their international collaboration.

\end{acknowledgements}

\newpage

    \begin{figure}
        \includegraphics[width=12cm]{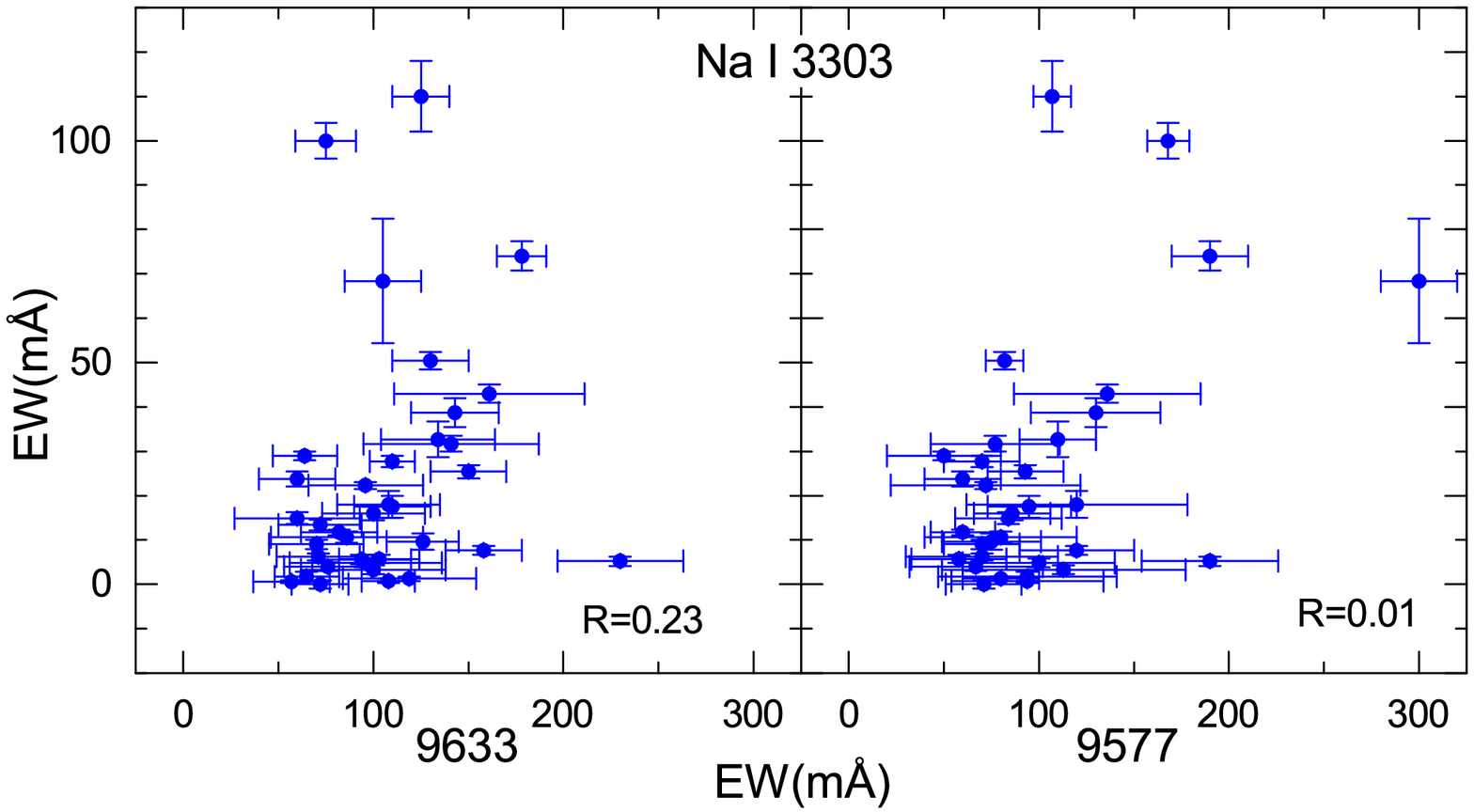}
        \caption{The same as Fig. \ref{H1} but for Na{\sc i}. }
        \label{Na1}
    \end{figure}

    \begin{figure}
        \includegraphics[width=12cm]{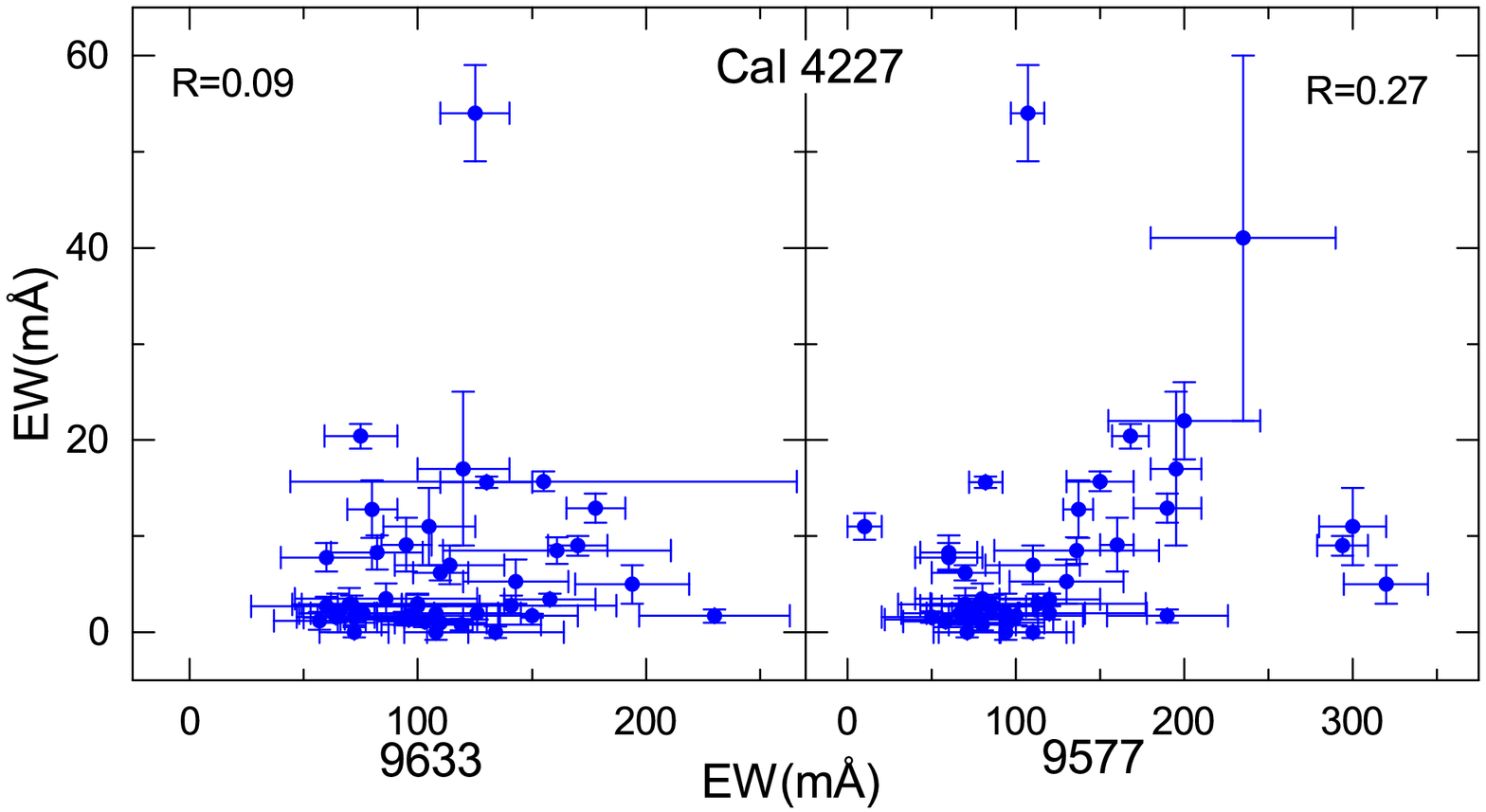}
        \caption{The same as Fig. \ref{H1} but for Ca{\sc i}. }
        \label{Ca1}
    \end{figure}

    \begin{figure}
        \includegraphics[width=12cm]{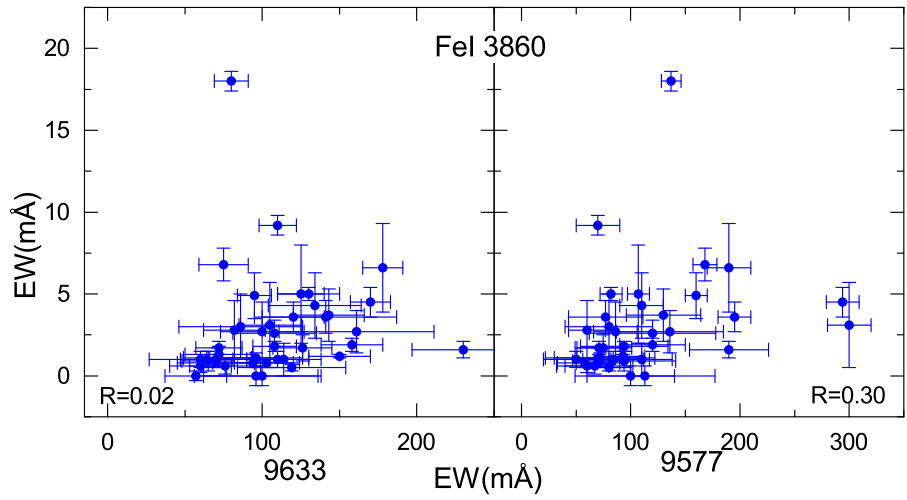}
        \caption{The same as Fig. \ref{H1} but for Fe{\sc i}. }
        \label{Fe1}
    \end{figure}

    \begin{figure}
        \includegraphics[width=12cm]{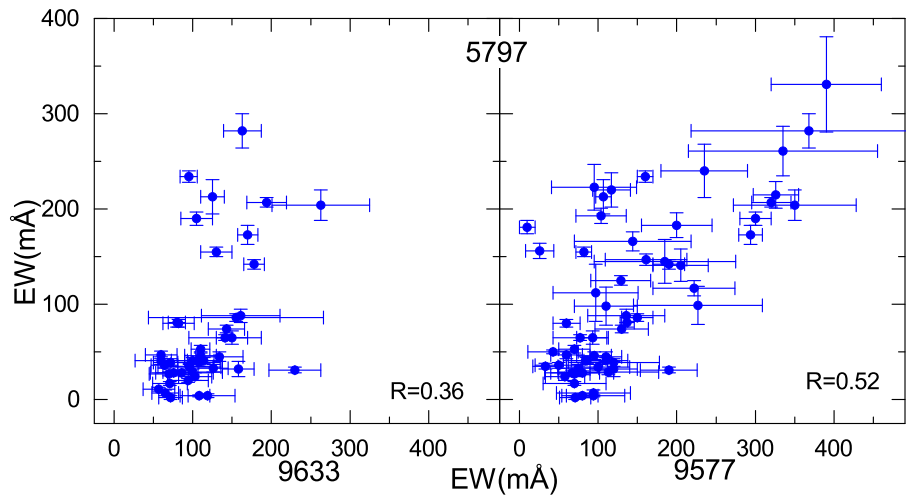}
        \caption{The same as Fig. \ref{6196} but for DIB 5797. }
        \label{5797}
    \end{figure}
    \begin{figure}
        \includegraphics[width=12cm]{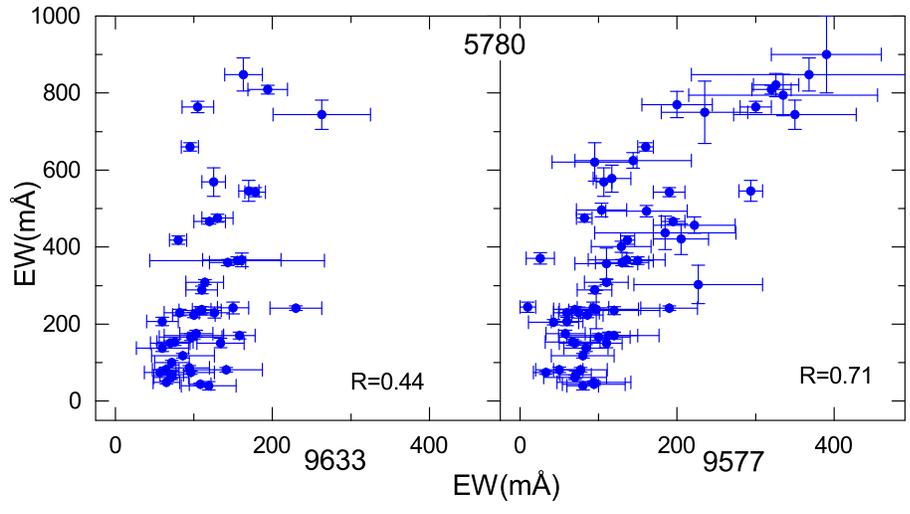}
        \caption{The same as Fig. \ref{6196} but for DIB 5780. }
        \label{5780}
    \end{figure}
    \begin{figure}
        \includegraphics[width=12cm]{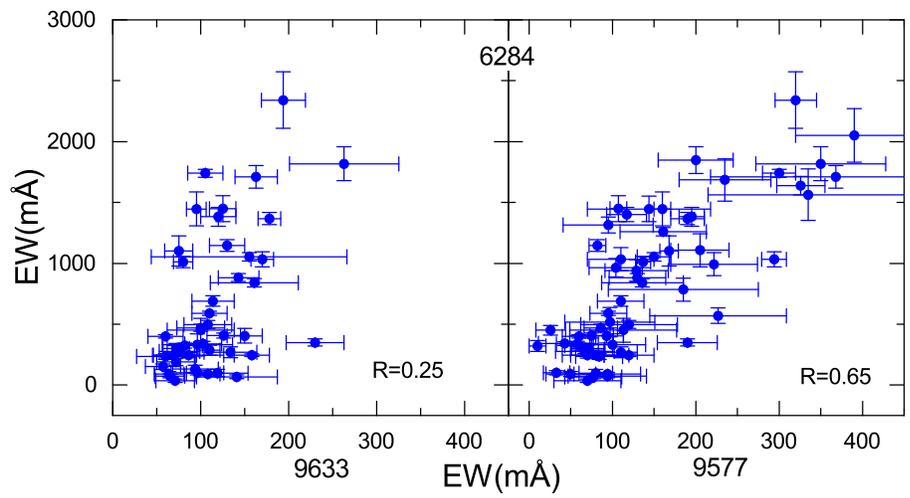}
        \caption{The same as Fig. \ref{6196} but for DIB 6284. }
        \label{6284}
    \end{figure}
    \begin{figure}
        \includegraphics[width=12cm]{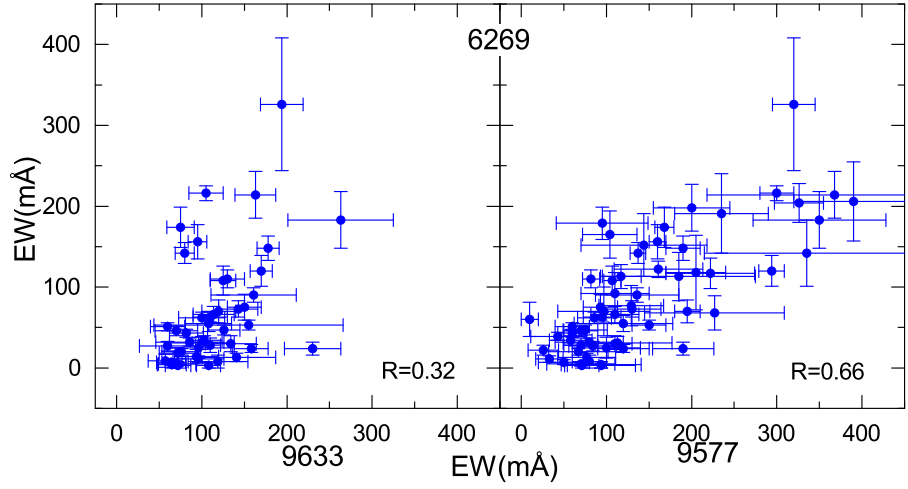}
        \caption{The same as Fig. \ref{6196} but for DIB 6269. }
        \label{6269}
    \end{figure}
    \begin{figure}
        \includegraphics[width=12cm]{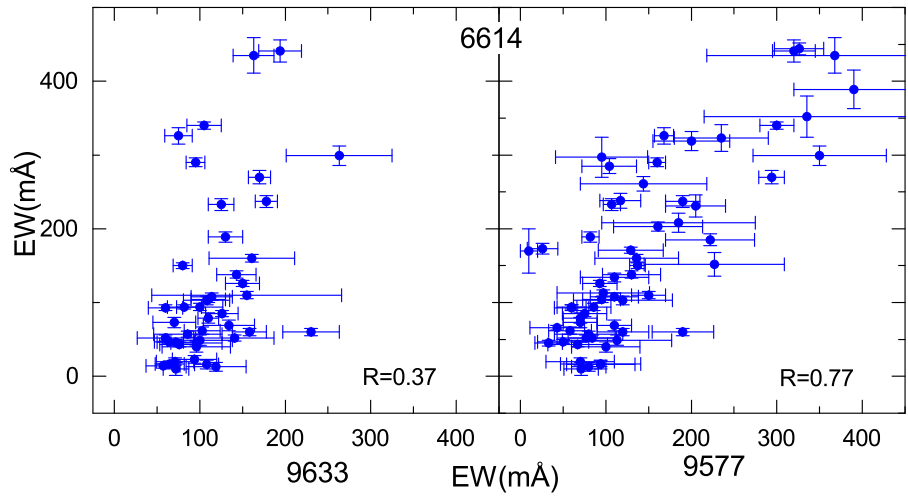}
        \caption{The same as Fig. \ref{6196} but for DIB 6614. }
        \label{6614}
    \end{figure}
    \begin{figure}
        \includegraphics[width=12cm]{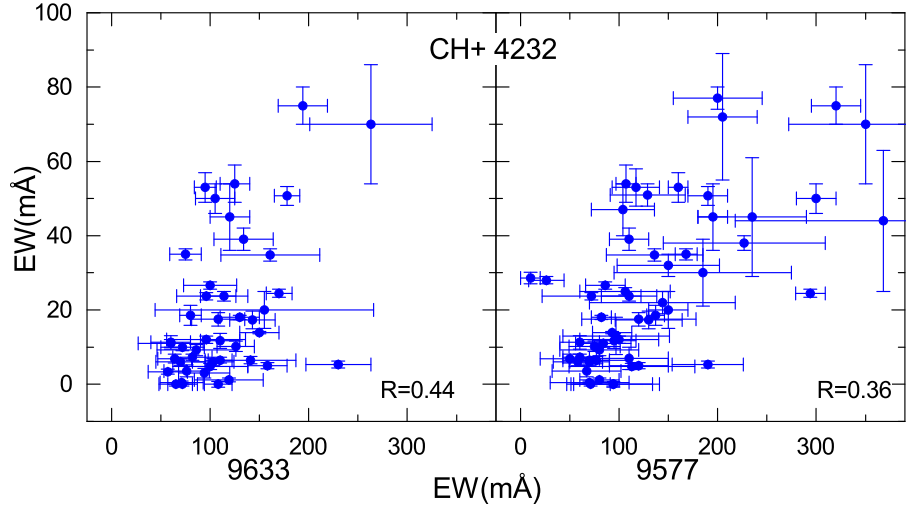}
        \caption{The same as Fig. \ref{CH4300} but for CH$^+$ 4232 \AA.}
        \label{CHp}
    \end{figure}
    \begin{figure}
        \includegraphics[width=12cm]{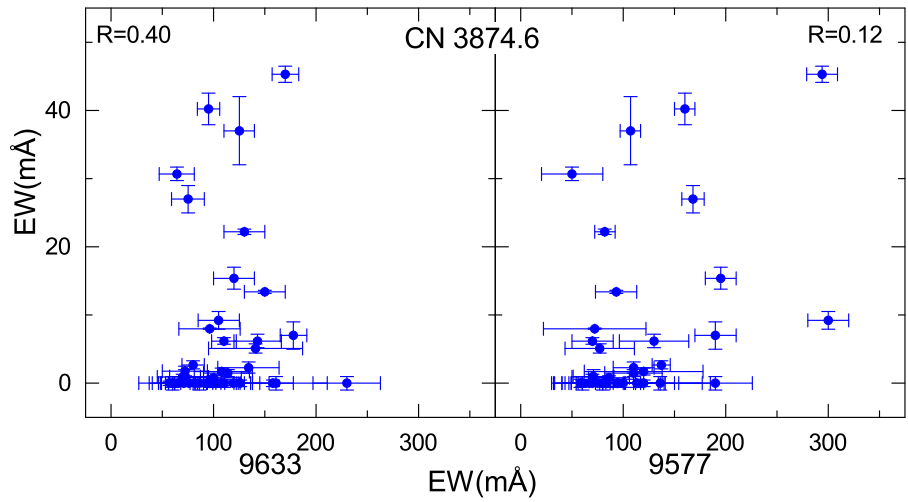}
        \caption{The same as Fig. \ref{CH4300} but for CN 3874.607 \AA.}
        \label{CN}
    \end{figure}

\end{document}